\documentclass[twocolumn]{aa}
\usepackage{graphicx}
\usepackage{txfonts}
\usepackage{natbib}

\newcommand{\req}[1]{(\ref{#1})}

\begin{document}

  \title{The Cellular Burning Regime in Type Ia Supernova Explosions}

  \subtitle{II.~Flame Propagation into Vortical Fuel}


  \titlerunning{Cellular Burning in SNe Ia---Flame
                Propagation into Vortical Fuel}

  \author{F. K. R{\"o}pke\inst{1},  W. Hillebrandt\inst{2}, \and
          J. C. Niemeyer\inst{3}}

  \authorrunning{F.~K.~R{\"o}pke et al.}

   \offprints{F. K. R{\"o}pke}

   \institute{Max-Planck-Institut f\"ur Astrophysik,
              Karl-Schwarzschild-Str. 1, D-85741 Garching, Germany\\
              \email{fritz@mpa-garching.mpg.de}
              \and
              Max-Planck-Institut f\"ur Astrophysik,
              Karl-Schwarzschild-Str. 1, D-85741 Garching, Germany\\
              \email{wfh@mpa-garching.mpg.de}
              \and
              Universit\"at W\"urzburg,
              Am Hubland, D-97074 W\"urzburg, Germany \\
              \email{niemeyer@astro.uni-wuerzburg.de}
             }

  \abstract{We investigate the interaction of thermonuclear flames in Type Ia
    supernova explosions with vortical flows by means of numerical
    simulations. In our study, we focus on
    small scales, where the flame propagation is no longer dominated by the
    turbulent cascade originating from large-scale effects. Here,
    the flame propagation proceeds in the cellular burning regime,
    resulting from a balance between the Landau-Darrieus instability and
    its nonlinear stabilization \citep{roepke2003a,roepke2003b}. The
    interaction of a cellularly
    stabilized flame front with a vortical fuel flow is explored applying
    a variety of fuel densities and strengths of the velocity
    fluctuations. We find that the vortical flow can break up the cellular
    flame structure if it is sufficiently strong. In this case the flame
    structure adapts to the imprinted flow field. The transition from
    the cellularly stabilized front to the flame structure dominated by
    vortices of the flow proceeds in a smooth way. The implications of the
    results of our simulations for Type Ia Supernova explosion models are
    discussed.

  }

  \maketitle

\section{Introduction}
\label{intro_sec}

In the canonical astrophysical model Type Ia supernovae (SNe Ia
in the following) are associated with thermonuclear explosions of white
dwarf (WD) stars \citep{hoyle1960a}. These WDs are composed of carbon
and oxygen. Due to the high temperature sensitivity of
the activation energy of carbon and oxygen burning, the reaction is
confined to a very narrow region and propagates as a combustion
wave. This defines a \emph{thermonuclear flame}. The theoretical
understanding of flame propagation in WDs provides the key to SN Ia
explosion models with the effective flame velocity being
the most important parameter. What are the phenomena that mediate the
combustion
wave? The answer to that question is given by the hydrodynamics of
combustion. Conservation laws across the flame admit two fundamental
modes of flame propagation: the \emph{detonation}, in which the flame
propagates due to shock waves, and the \emph{deflagration}, which is
based on microphysical transport phenomena such as heat conduction and species
diffusion. 

Flame propagation in the detonation mode is certainly
easiest to model. Here the flame propagation velocity with respect
to the unburnt material is larger than the corresponding sound
speed. Therefore the WD star has no
time to expand prior to the incineration in this \emph{detonation model} of
SN Ia explosions and all the material is consumed at
high densities resulting in ashes consisting solely of iron-peak elements
\citep{arnett1969a}. However, this is in conflict with
observational spectra, which show strong indication of intermediate
mass elements.

Consequently, the star has to expand considerably before it is
processed by the thermonuclear flame. This is
possible if flame propagation starts out in the deflagration mode,
since here the microphysical transport processes lead to a subsonic flame
velocity. In the present study, we refer to that \emph{deflagration
  model} of SNe Ia. 
Moreover, underlying to our model is
the so-called single-degenerate scenario, where the progenitor is a
binary system with the WD accreting matter from the
non-degenerate companion until it reaches the Chandrasekhar mass. At
this point, thermonuclear reaction ignites near the center of the WD
and propagates outward as a deflagration flame. For a review of SN Ia
explosion models we refer to \cite{hillebrandt2000a}.

In the framework of the deflagration model, recent large-scale SN Ia explosion
models have been
quite successful \citep{reinecke2002d, gamezo2003a}. Nevertheless, the determination
of the flame propagation speed is much harder in this case. A planar
deflagration flame would be far too slow to cause powerful SN Ia
explosions, but these models predict an acceleration of the
deflagration flame due to interaction with
turbulence. The resulting energy release is sufficient to unbind the
WD star. Following the reasoning by \citet{niemeyer1997b}, the turbulent
motions are evoked by large-scale instabilities---such as
the Rayleigh-Taylor (buoyancy) instability and the Kelvin-Helmholtz
(shear) instability---and decay to smaller
scales forming a turbulent eddy cascade
\citep{richardson1922a}. Eddies of that cascade wrinkle the flame and
increase its surface. This corresponds to an acceleration of the
effective flame propagation speed. However, the vast range of spatial
scales relevant to the problem of SN Ia explosions (from the radius of
the star down to the width of a typical flame it covers about 11 orders
of magnitude) restricts these models to the simulation of effects on
the largest scales only. This makes it impossible to directly
determine the turbulent flame propagation velocity in the simulations and
therefore assumptions on the physics on
unresolved small scales must be made. The fundamental idea here is
that flame propagation is dominated by the turbulent cascade that
is driven solely from large-scale instabilities and that no other
effects contribute to the generation of turbulence. 

However, due to
the scaling law of the turbulent cascade, which predicts the turbulent
velocity fluctuations to monotonically decrease with smaller scale,
there exists a certain length scale below which the flame burns faster
through the corresponding eddies than these can deform it. For chemical flames, this
scale is identified with the Gibson scale \citep{peters1986a} and we
will use this term in the context of thermonuclear combustion,
too. Below the Gibson scale the flame propagates in a ``frozen
turbulence field'' and is no longer affected by the eddy
cascade. Large scale models assume stable flame propagation
here. Although some theoretical concepts support this idea, no
hydrodynamical simulations provided convincing proofs yet.

In order to fill
that gap in SN Ia models we performed numerical studies of flame propagation
around the Gibson scale. The basic physical effects that determine
flame propagation here are the hydrodynamical \emph{Landau-Darrieus
instability} \citep{darrieus1938a,landau1944a} and the counteracting
nonlinear stabilization of the flame \citep{zeldovich1966a}. The
balance of the two effects gives rise to a stable cellular flame
structure characterizing the \emph{cellular burning regime}.
The goals
of our investigations were to determine whether or not
the cellular burning regime exists for thermonuclear flames in WD matter
and to test how robust the anticipated cellular stabilization is under varying
conditions. On the one hand, these studies examine a
fundamental assumption of present SN Ia explosion models and on the
other hand they are intended to
explore the possibility of new physical effects that may have to be
included into these. In particular, the question of the
robustness of the cellular stabilization of the flame is closely
connected to the so-called \emph{delayed detonation models} of SN Ia
explosions. In these models, the flame starts out as a deflagration
wave, but later turns into a supersonic detonation. Empirical
one-dimensional SN Ia models support this idea, since those lead to
quite realistic spectra if a \emph{deflagration-to-detonation transition}
(DDT) is imposed artificially at low fuel densities
\citep{hoeflich1996a,iwamoto1999a}. Although DDTs are frequently
observed in terrestrial combustion, a mechanism that could
account for such a transition under conditions of SN Ia explosions
could not be identified so far \citep{niemeyer1999a}. However, one
possibility is a breakdown
of the flame stabilization and a subsequent generation
of turbulence on small scales by the flame itself, that would result
in an additional acceleration of the flame. This \emph{active
  turbulent combustion} (ATC) mechanism has been proposed by
\citet{niemeyer1997b} on the basis of a work by
\citet{kerstein1996a}. \cite{niemeyer1997b} suggest that the
cellular flame stabilization could break down at the Gibson scale
leading to ATC. One motivation for our study is to examine this
idea.

The present paper is the third in a series of publications on the
investigation of the cellular burning regime in Type Ia supernova
explosions. After the introduction of the numerical methods to study
this regime together with providing
some examples of application \citep{roepke2003a}, the implementation was used
to study flame propagation into quiescent fuel \citep{roepke2003b}. The results
obtained there (to be summarized in the next section) raised the
question of the interaction of a cellular flame with turbulent
flows. This topic is also motivated from the supernova model. Around
the Gibson scale one can still expect turbulent velocity fluctuations
of the order of the laminar burning speed of the flame originating
from the turbulent cascade. Additionally, pre-ignition convection in
the WD may result in relic turbulent motions of considerable
strength. \cite{hoeflich2002a} claim surprisingly high values of
convective velocities.

A brief description of the physical effects underlying the cellular
burning regime and of the numerical methods applied to model it will
be provided in the next section. In Sect.~\ref{vorf_sec} the
results of our numerical simulations will be presented and
discussed. Finally, we will draw conclusions emphasizing the
implications of our results for SN Ia models. 

\section{The cellular burning regime and its numerical simulation}

Since our study aims at the flame evolution around the Gibson scale, which
is expected to be well-separated from the scale of the flame width for
fuel densities above $10^7 \,\mathrm{g} \,\mathrm{cm}^{-3}$, we will treat the
flame in the \emph{discontinuity approximation} in the following. That
is, we describe the flame as a simple discontinuity between fuel and
ashes and ignore its internal structure. Note that this picture does
not account for the microphysical transport phenomena and therefore
the laminar burning speed of the flame (i.e.~the velocity of a planar
deflagration flame) is not intrinsically given in this
model. 

The Gibson scale is determined by this laminar burning speed $s_l$, by
the velocity fluctuations at the integral scale of turbulence, and
by the scaling law of the eddy cascade. 
\citet{timmes1992a} determined the laminar burning velocity by means of
one-dimensional simulations fully resolving the internal structure of
the flame, and we apply
the fitting formula given there to obtain the value for a particular
composition and density of the fuel. As has been mentioned by \citet{roepke2003b},
this formula is afflicted with some uncertainties, which
introduce a considerable ambiguity to the exact value of the Gibson
scale. Following the approach of \citet{roepke2003b}, we will
arrange our simulations
at length scales around $10^4 \,\mathrm{cm}$ as a crude estimate of the
Gibson length, since our goal is to
qualitatively explore the possible effects here rather than to perform
highly precise measurements.

Below the Gibson scale, flame propagation is determined by the
competition between the Landau-Darrieus (LD)
instability and its nonlinear stabilization. The LD instability is of
pure hydrodynamical origin and results from the density contrast
across the flame front in combination with mass flux
conservation. This leads to a refraction of the stream lines at the
flame causing a flow field that enhances initial perturbations of the
planar flame shape. The fact that the LD instability acts on
thermonuclear flames under conditions of SN Ia
explosions has been proven by means of a full hydrodynamical
simulation by \citet{niemeyer1995a}. The refined numerical model that
was described by \citet{roepke2003a} allowed us to investigate this
instability in detail. We simulated the propagation of a sinusoidally
perturbed flame front into quiescent fuel. The initial linear regime
of flame evolution was shown to be indeed determined 
by the LD instability. The flow field
accounting for this effect was
apparent in our simulations \citep{roepke2003b}. Moreover, the growth
rate of the amplitude of the perturbation was found to be
consistent with the dispersion relation that \citet{landau1944a}
derived by means of a linear stability analysis.

The nonlinear stabilization of the flame is a geometrical
effect. \citet{zeldovich1966a} explained the mechanism by following
the propagation of a perturbed flame by means of Huygens' principle of
geometrical optics (see also Fig.~1 in \citealt{roepke2003a}). Once the
perturbation has grown to a critical size, former recesses of the
flame front  develop into cusps between bulges of the front. It
can easily be shown that the propagation velocity $v_\mathrm{cusp}$ of
such a cusp exceeds the laminar burning velocity $s_l$ of the other
parts of
the flame by
\begin{equation}
v_\mathrm{cusp}= \frac{s_l}{\cos \theta},
\end{equation}  
where $\theta$ denotes the inclination angle of the bulges adjacent
to the cusp. This effect balances further growth of the perturbation
due to the LD instability and causes a cellular shape of the flame
front. Therefore the regime of flame propagation at scales where it is
dominated by the interplay between the LD instability and its nonlinear
stabilization will 
be termed \emph{cellular burning regime} in the following. 

The nonlinear stage of flame evolution could not be reached by early
attempts of hydrodynamical simulations of flame propagation
\citep{niemeyer1995a}. However, applying a semi-analytical description
of flame evolution, \citet{blinnikov1996a} explored effects of the
cellular burning regime in the context of SN Ia explosions and also
\citet{bychkov1995a} discussed the cellular flame stabilization.
With help of the numerical
implementation that was described by \citet{roepke2003a} on the basis
of the work by \citet{reinecke1999a}, it was possible to prove by means
of a hydrodynamical simulation, that
cellular stabilization holds for thermonuclear flames in SN Ia
explosions \citep{roepke2003a,roepke2003b}. In connection with
flame interaction with vortical flows, which will be studied in the
following, the most important results of
the  investigation of
the cellular stabilization of the flame propagating into \emph{quiescent}
fuel were the following:
\begin{itemize}
\item The flame stabilizes in a cellular shape for fuel densities down
  to $10^7 \,\mathrm{g} \,\mathrm{cm}^{-3}$. However, at the lowest fuel
  densities, the stabilization can only be revealed with high
  numerical resolution. This may be interpreted as a hint to increased
  sensitivity of the cellular flame stabilization to (numerical) noise
  at low fuel densities.
\item The evolution of the flame shape tends to a steady state
  consisting of only one 
  cell filling the entire computational domain. This is consistent with
  results from semi-analytical and analytical studies based on the
  Sivashinsky-equation \citep{gutman1990a,thual1985a}. The single-cell
  solution establishes independently of the wavelength of the initial flame
  perturbation via merging of small cells.
\item Applying a simulation setup of an on average planar flame
  configuration in a finite computational domain leads to a dependence
  of the alignment of the finally emerging domain-filling single-cell
  structure on the boundary conditions imposed transverse to the
  direction of flame propagation. In contrast to periodic boundaries,
  the crest will align in the center for reflecting boundary
  conditions. This is consistent with
  semi-analytical results \citep{gutman1990a}. 
\item The emerging cellular shape of the flame leads to an increased
  flame surface and equivalently an increased effective flame
  propagation speed as compared to a planar flame front.
\item Depending on the width of the computational domain, the
  fundamental single-cell structure may be superimposed by a
  short-wavelength cellular pattern. However, these small cells are
  advected towards the cusp where they disappear. This mechanism is
  consistent with theoretical predictions by \citet{zeldovich1980a}.
\end{itemize} 
These results will be helpful in the interpretation of the simulations
of flame propagation in \emph{vortical} flows. The possibility of a weaker cellular
stabilization at low fuel densities is a further motivation to test
the interaction of the flame with physical noise, represented by a vortical flow
field.

We will only briefly outline the numerical
implementation here and refer to \citet{roepke2003a} and
\citet{reinecke1999a} for detailed
discussions. The description of the flame in the discontinuity
approximation allows us to separate the flame propagation from the
hydrodynamics in a first step. 
The hydrodynamics part of the numerical
implementation thus reduces to the solution of the reactive Euler
equations, which is performed applying a finite-volume approach. In
particular we employ the piecewise parabolic method (PPM) suggested by
\cite{colella1984a} in the \textsc{Prometheus} implementation by
\cite{fryxell1989a}. 
Flame propagation is modeled using the level-set method
(e.g.~\citealt{osher1988a, sussman1994a, smiljanovski1997a,reinecke1999a}). For the correct reproduction of
hydrodynamical effects, such as the LD instability, a precise coupling
between flame and flow turned out to be essential
\citep{roepke2003a}. This is provided by applying the in-cell
reconstruction/flux-splitting technique introduced by
\citet{smiljanovski1997a}.

\section{Flame interaction with a vortical flow}
\label{vorf_sec}

In the following we will present the results from simulations of the
interaction of cellular flames with vortical flows. After discussing
the simulation setup we will make some general remarks on what can be
expected in the simulations. This will be followed by a survey of
flame evolution depending on the strength of the vortices for an
exemplary case at a fixed fuel density. Finally, we will capture the
simulation results for various fuel densities in a quantitative way.

\subsection{Simulation setup}
\label{vorf_setup_sec}

\begin{figure}[t]
\centerline{
\includegraphics[width=0.95 \hsize, viewport = 0 0 310 284,clip]{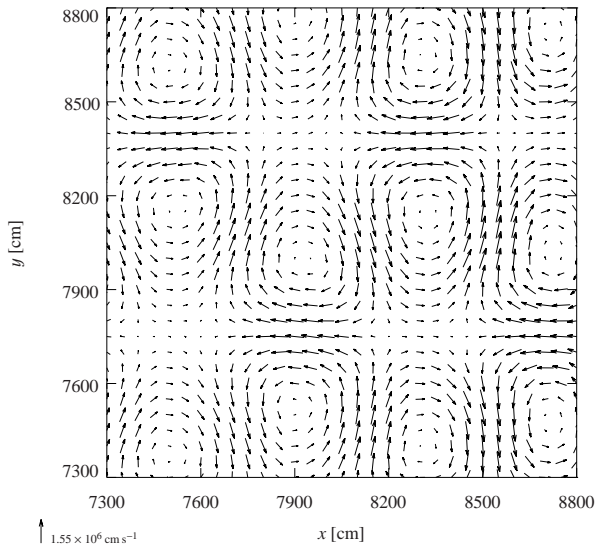}}
\caption{Vortical flow field as applied in the simulations (here for
  the case $\rho = 5 \times 10^8 \,\mathrm{g} \,\mathrm{cm}^{-3}$). The frame
  of reference is comoving with the vortices. The snapshot is taken
  after 1.5 flame crossing times over the entire computational
  domain. \label{vort_fig}}
\end{figure}

The basic setup of our simulations was similar to that used to describe
flame propagation into quiescent fuel
\citep{roepke2003a,roepke2003b}. The flame was initialized in  a
computational domain
with periodic boundary conditions transverse to the direction of flame
propagation.
In order to capture the full flame front in
cases of strong deformation, we changed from a quadratic domain to a
rectangle of $300 \times 200$ computational cells. The cell width was
set to $\Delta x = \Delta y = 50 \, \mathrm{cm}$. Several experiments
with initially planar flames yielded drastic responses of the flame
shape when the vortices encountered it. This can be attributed to the
high sensitivity of planar flames to perturbations. Consequently, in
order to make
the results of different simulations comparable and to enable a
quantification of the flame evolution, the flame was perturbed
initially in a sinusoidal way with eight periods fitting into the
domain. This initial perturbation grows due to the LD instability and
stabilizes in a cellular pattern, before the injected vortices
reach the flame front. In this way it was possible to study the
interaction of a stabilized flame with a vortical flow field
preventing the incoming flow from unpredictably deforming the flame
shape at the first encounter.

\begin{figure}[t]
\centerline{
\includegraphics[width=0.9 \hsize, viewport=-10 0 259 508,clip]{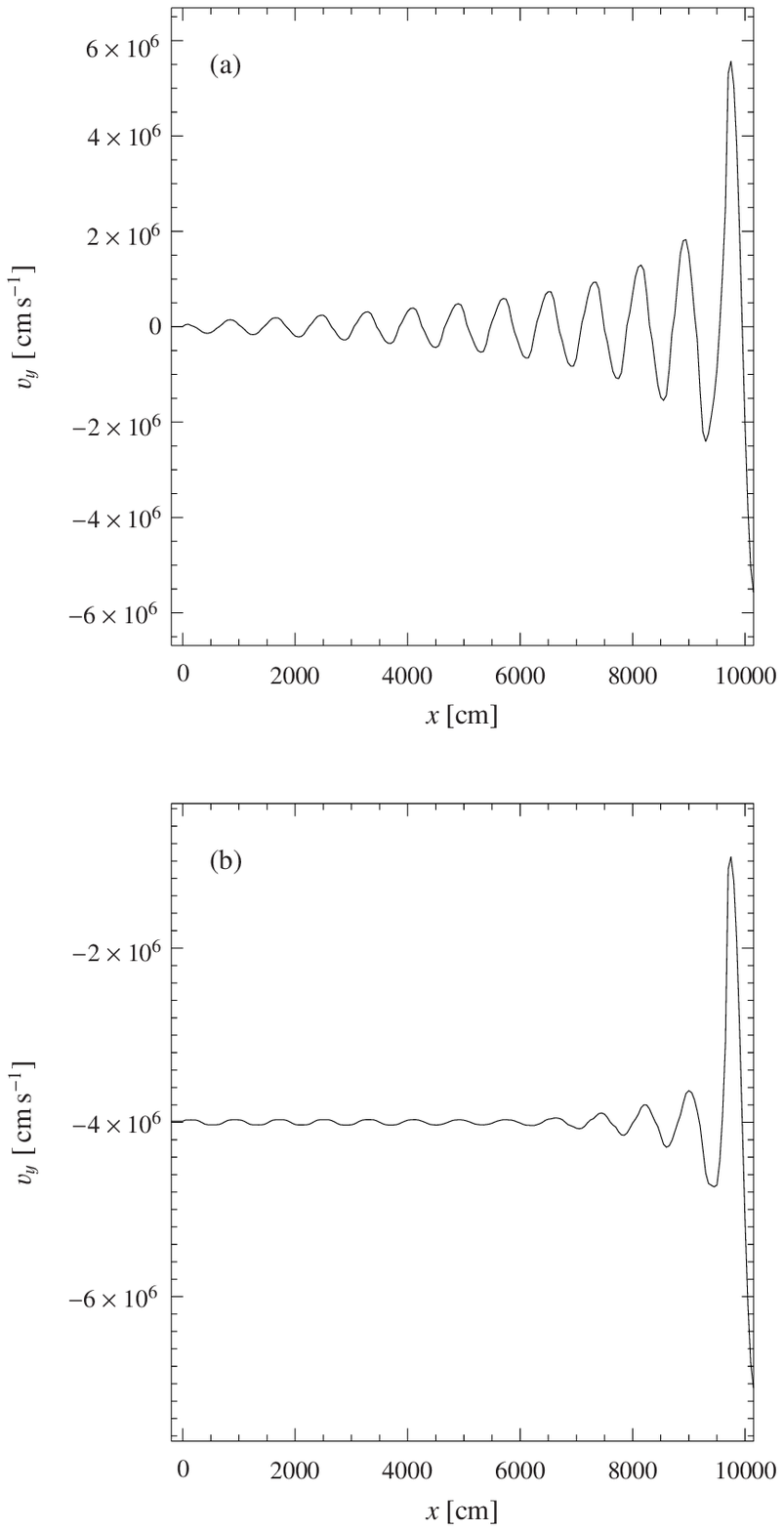}}
\caption{Profile of the $y$-component of the vortical flow field
  parallel to the $x$-axis \textbf{(a)} $\rho=5 \times 10^8 \,\mathrm{g}
  \,\mathrm{cm}^{-3}$; \textbf{(b)} $\rho=1 \times 10^8 \,\mathrm{g}
  \,\mathrm{cm}^{-3}$. The snapshots are taken
  after 1.5 flame crossing times over the entire computational domain.\label{vely_fig}}
\end{figure}

The numerical investigation of flame interaction with turbulence is
generally an 
intricate issue. In principle, the straight forward way to go would be
to produce an
isotropic turbulence field by external forcing and to set up a flame in
this field. This approach is, however, too expensive for the purpose
of the parameter study we are aiming at. 
For this reason we simply modified the inflow boundary condition in
order to inject a vortical flow field instead of quiescent fuel. The
flow field is not stirred  and turbulence is
not actively produced inside the computational domain by external
forcing. As a consequence turbulence injected here  partly decays
before reaching the flame front---an effect which we have to take into
account in the measurement of characteristic quantities.

\begin{table*}
\caption{Setup values for the simulations of the flame
  evolution.
\label{values_tab}}
$$
\begin{array}{p{0.11\linewidth}p{0.12\linewidth}p{0.12\linewidth}p{0.08\linewidth}
              p{0.08\linewidth}p{0.12\linewidth}p{0.12\linewidth}p{0.12\linewidth}}
\hline\hline
\noalign{\smallskip}
Label & \mbox{$\rho_u [\,\mathrm{g} \,\mathrm{cm}^{-3}]$} & \mbox{$\rho_b [\,\mathrm{g}
\,\mathrm{cm}^{-3}]$} & \mbox{$\mu$} & \mbox{$\mathit{At}$} & \mbox{$s_l [\,\mathrm{cm} \,\mathrm{s}^{-1}]$} (TW) &
\mbox{$e_{i,u} [\,\mathrm{erg} \,\mathrm{g}^{-1}]$}
& \mbox{$e_{i,b} [\,\mathrm{erg} \,\mathrm{g}^{-1}]$}\\
\noalign{\smallskip}
\hline
\noalign{\smallskip}
\mbox{$1.25 \times 10^7$} & \mbox{$ 1.232 \times 10^7$} & \mbox{$3.830 \times 10^6 $} & 3.22 &
0.526 & \mbox{$2.86 \times 10^5$} & \mbox{$2.005 \times 10^{17}$} & \mbox{$8.99 \times
10^{17}$}\\
\mbox{$2.5 \times 10^7$} & \mbox{$ 2.484 \times 10^7$} & \mbox{$8.69 \times 10^6 $} & 2.86 &
0.482 & \mbox{$5.01 \times 10^5$} & \mbox{$2.56 \times 10^{17}$} & \mbox{$9.50 \times
10^{17}$}\\
\mbox{$5 \times 10^7$} & \mbox{$ 4.988 \times 10^7$} & \mbox{$2.071 \times 10^7 $} & 2.41 &
0.413 & \mbox{$8.74 \times 10^5$} & \mbox{$3.584 \times 10^{17}$} & \mbox{$1.051 \times
10^{18}$}\\
\mbox{$7.5 \times 10^7$} & \mbox{$ 7.50 \times 10^7$} & \mbox{$3.345 \times 10^7 $} & 2.24 &
0.383 & \mbox{$1.21 \times 10^6$} & \mbox{$4.29 \times 10^{17}$} & \mbox{$1.13 \times
10^{18}$}\\
\noalign{\smallskip}
\hline
\end{array}
$$
\end{table*}

As suggested by \citet{helenbrook1999a}, we applied an
\emph{oscillating inflow boundary condition} on the right hand side of the
computational domain, which generates a vortical
velocity field approaching the flame. The velocity at the boundary now
reads:
\begin{eqnarray}
v_x &=& s_l \left\{ -1 + v' \sin 2 k \pi y \cos 2 k \pi (x - t
  s_l)\right\} \label{osc_bc_1}\\
v_y &=& s_l\, v' \cos 2 k \pi y \sin 2 k \pi (x - t
  s_l). \label{osc_bc_2}
\end{eqnarray}
The parameter $v'$ characterizes the strength of the imprinted
velocity fluctuations and $k$ denotes the wavenumber of the
oscillation.
This produces what is termed ``square vortices'' by
\citet{helenbrook1999a}. A similar flow field is applied by
\citet{vladimirova2003a} who name it ``cellular flow'' and also
\citet{zhu1994a} use a flow with cellular vortices to study flame
propagation in it. The particular flow has the
advantage of being very simple---ideally containing only one Fourier
mode. This helps to make the effects visible more clearly and
simplifies the interpretation.
Figure \ref{vort_fig} illustrates the vortical
flow applied in our simulations. It shows an example at a density of
$\rho= 5 \times 10^8 \,\mathrm{g} \,\mathrm{cm}^{-3}$. In Fig.~\ref{vely_fig}a the
$y$-component
of the velocity is plotted against the $x$-coordinate for the same
simulation. Obviously, the velocity fluctuation is
damped while propagating from the inflow boundary into
the computational domain. The damping is even stronger for lower
densities. Figure~\ref{vely_fig}b shows a plot for $\rho = 1\times 10^8
\,\mathrm{g} \,\mathrm{cm}^{-3}$.

\begin{figure}[h!]
\centerline{
\includegraphics[width = 0.9735 \hsize, viewport = 0 0 275 540]{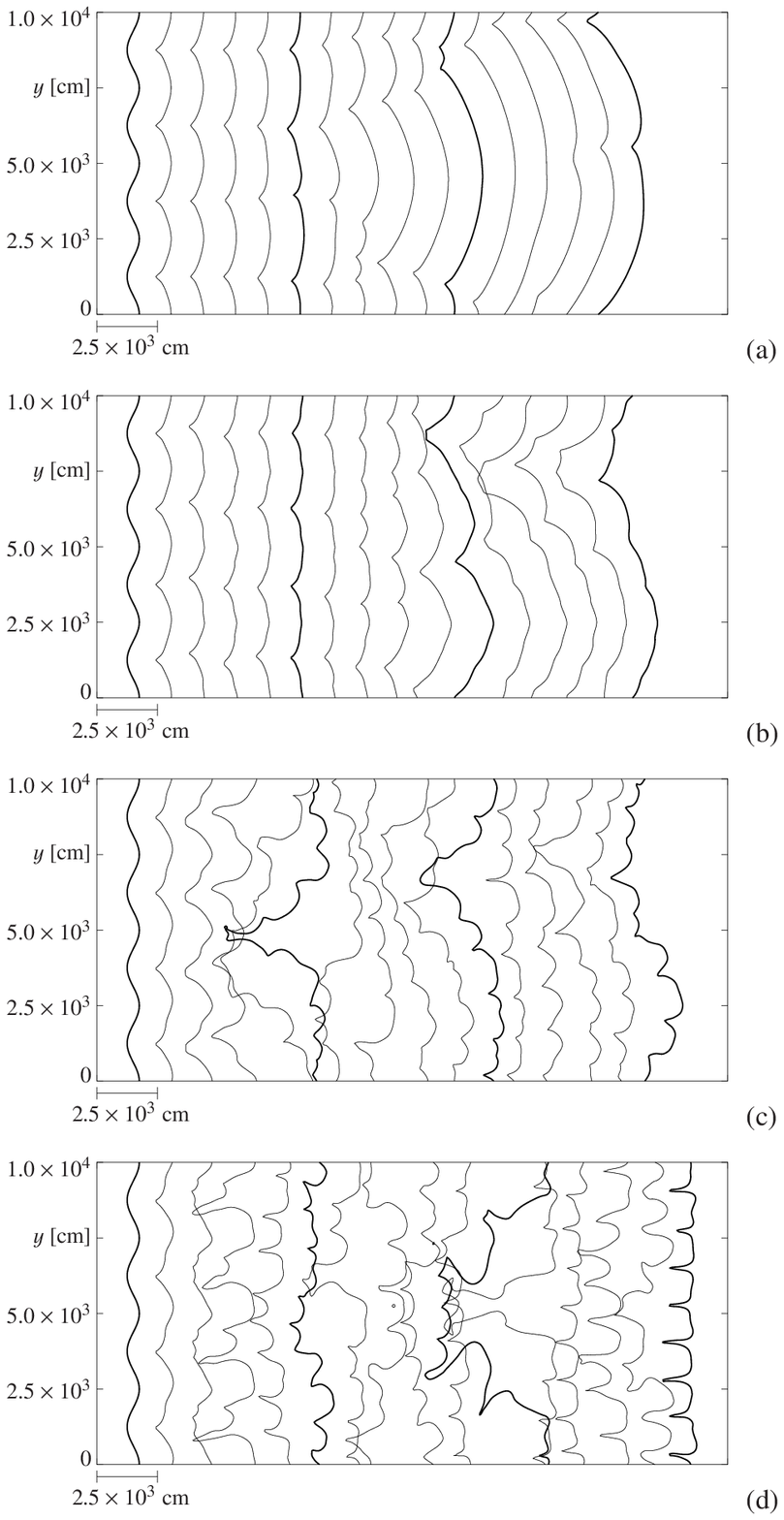}}
\caption{\textbf{(a)} Flame  propagation into quiescent fuel at a density of $5 \times 10^7
  \,\mathrm{g} \,\mathrm{cm}^{-3}$ and interaction with a vortical fuel field for velocity
  fluctuations at the right boundary of \textbf{(b)} $v'/s_l = 0.7$,  \textbf{(c)}
  $v'/s_l = 2.0$, and \textbf{(d)} $v'/s_l = 2.5$. Each contour represents a
  time step of $3.2 \times 10^{-3} \,\mathrm{s}$. \label{evo_vort_fig}}
\end{figure}

\begin{figure*}[ht]
\centerline{
\includegraphics[width= 0.95 \textwidth]{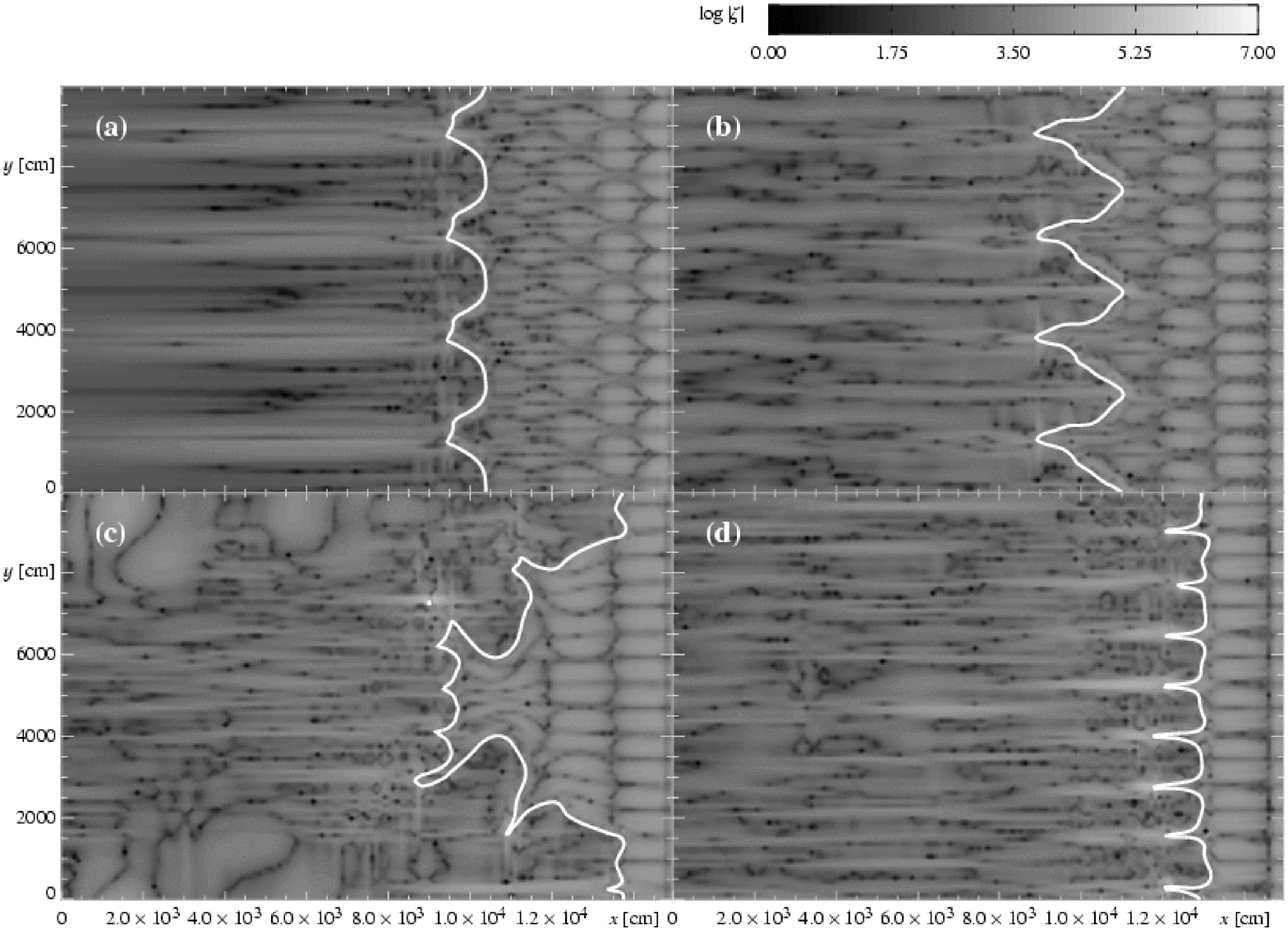}}
\caption{Flame evolution for a fuel density of $5
\times 10^7 \,\mathrm{g} \,\mathrm{cm}^{-3}$ and a velocity fluctuation of $v'/s_l
= 2.5$ at the right boundary. Snapshots taken at time steps
\textbf{(a)} $t=4.0 \times 10^{-3} \,\mathrm{s}$, 
\textbf{(b)} $t=8.0 \times 10^{-3} \,\mathrm{s}$, 
\textbf{(c)} $t=3.2 \times 10^{-2} \,\mathrm{s}$, and 
\textbf{(d)} $t=4.8 \times 10^{-2} \,\mathrm{s}$. The vorticity $\zeta$ (cf.~eq.~\req{vorticity_def})
  is color-coded and the flame position is indicated by solid
  white curves.
\label{panel_fig_3}}
\end{figure*}

\begin{figure*}[ht]
\centerline{
\includegraphics[width= 0.95 \textwidth]{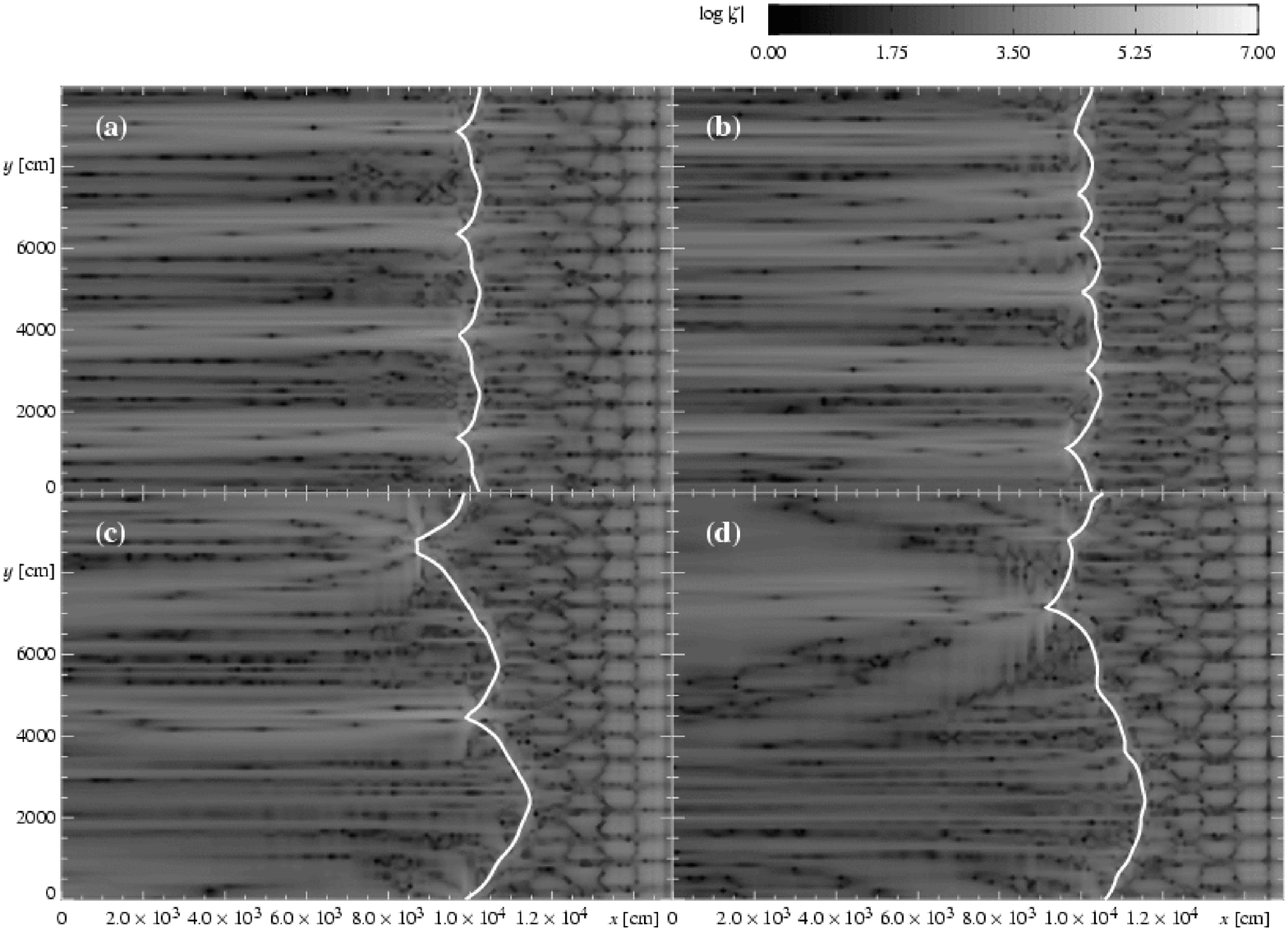}}
\caption{Flame evolution for a fuel density of $5
\times 10^7 \,\mathrm{g} \,\mathrm{cm}^{-3}$ and a velocity fluctuation of $v'/s_l
= 0.7$ at the right boundary. Snapshots taken at time steps 
\textbf{(a)} $t=8.0 \times 10^{-3} \,\mathrm{s}$, 
\textbf{(b)} $t=2.4 \times 10^{-2} \,\mathrm{s}$, 
\textbf{(c)} $t=3.2 \times 10^{-2} \,\mathrm{s}$, and 
\textbf{(d)} $t=4.8 \times 10^{-2} \,\mathrm{s}$. The vorticity $\zeta$ (cf.~eq.~\req{vorticity_def})
  is color-coded and the flame position is indicated by solid
  white curves.
\label{panel_fig_2}}
\end{figure*}

In order to be able to follow the long-term flame
evolution, the simulations are carried out in a frame of reference
comoving with the flame. 
For a planar flame this could be achieved by
applying an inflow condition on the right hand side of the domain,
where fuel enters at the laminar burning velocity, and an outflow
condition on the opposite side of the domain. Since flame wrinkling
and subsequent surface enhancement due to the LD instability and
interaction with turbulent eddies will accelerate the flame,
additional measures have to be taken in order to keep the flame centered in the
domain. In the simulations presented by
\citet{roepke2003a,roepke2003b} the grid was simply shifted according
to the flame motion
relative to $s_l$. However, this approach is not consistent with
the oscillating boundary condition. Therefore we modified
eq.~\req{osc_bc_1} to
\begin{equation}
v_x = s_l \left\{ -\frac{A}{A_\mathrm{planar}} + v' \sin 2 k \pi y \cos 2 k \pi (x - t
  s_l)\right\},
\end{equation}
where $A$ denotes the actual flame surface (measured in each time
step) and $A_\mathrm{planar}$ stands for the surface of the
corresponding planar flame, which is equivalent to the
width of the domain in our case. This ensures
that the velocity with which the fuel enters the domain adapts to the
current effective flame propagation speed (cf.~eq.~\req{v_eff_eq} in
Sect.~\ref{param_sec}) and the position of the
flame relative to the grid is kept approximately fixed.

We performed a number of simulations with different fuel densities. For
convenience of the reader, a compilation of the setup values is
given in Table \ref{values_tab}.

\subsection{What can be expected?}

Before we present the results of our numerical study we discuss
what we actually are looking for in the simulations. That is, we
somehow have to define how to discriminate between stable flame
propagation and a breakdown of the stabilization. Why is this a question
of definition? From the conjecture of \emph{active
turbulent combustion} it may appear that a breakdown of the stabilization
must be obvious from a rapid and unlimited growth of the flame
propagation velocity. However, this cannot be expected to
occur in a numerical simulation where the growth of the flame surface
due to ATC would be limited by discretization and
resolution. Hence, the maximum effect that can be anticipated is a rapid
increase in flame surface and burning velocity and a saturation at
some higher value. A second possibility is that with increasing
strength the vortical flow
starts to dominate the flame evolution.

On the other hand, how can we guarantee flame
stabilization? Owing to the high computational costs it will not be
possible to follow the flame propagation for an arbitrarily long
time. However, in case of stabilization we can actually make use of the
knowledge of the flame pattern that is to be expected in case of
stability in our specific simulation setup.
From the preceding studies of flame propagation into quiescent fuel
\citep{roepke2003a,roepke2003b} it
is known that the flame finally
stabilizes in a single  domain-filling cusp-like structure for our setup.
This structure may  be superimposed by a
smaller-scale cellular pattern in sufficiently resolved simulations. A
similar result can be anticipated for interaction with weak imprinted
vortices. 

Thus we may expect two extremal behaviors of the flame. In one case
the flame stabilizes and the incoming perturbation fails to break up the
cellular pattern. The cells of the stabilized small wavelength pattern
caused by the initial perturbation of the flame front will then merge
forming the single domain-filling cusp which then propagates
stably. Alternatively, if the intensity of the incoming
vortices is high enough to destroy the stabilization, no single
domain-filling cusp will finally emerge. In this second case the flame
should rather show a transient pattern.

\subsection{General features}
\label{vort_res_sec}

As an example, we consider the case of a fuel density amounting to $5
\times 10^7 \,\mathrm{g} \,\mathrm{cm}^{-3}$ (cf.~Table
\ref{values_tab}). Figure \ref{evo_vort_fig} gives an overview over
the flame evolution in interaction with vortical fuel flows of
different strengths. The values $v'/s_l = 0.7$, $v'/s_l = 2.0$, and
$v'/s_l = 2.5$ indicate the amplitudes of velocity fluctuations
imposed at the right boundary of the domain. Note, however, that they
do not necessarily represent the values experienced by the flame front
for reasons given in section \ref{vorf_setup_sec}.

Figure
\ref{evo_vort_fig}a provides the comparison with flame propagation
into quiescent fuel for the chosen setup. In this simulation $v'$
was set to zero.
It resembles the features of the simulations presented by
\citet{roepke2003a,roepke2003b}.
Due to the LD instability,
the initial perturbation grows and the flame shape evolves into a
cellular pattern in the nonlinear regime. 
The subsequent contours show the
``merging'' of the short-wavelength cells imprinted by the
initial condition resulting in the formation of larger cells. 
This behavior was observed in a simulation of the long-term flame
evolution in quiescent fuel \citep{roepke2003b}. There a
cusp-like structure that was centered in the domain finally emerged as
the steady state structure. The
tendency to form a domain-filling cell is also apparent in
Fig.~\ref{evo_vort_fig}a. However, following
the evolution until the steady-state is reached is very expensive and
shall not be tried here.
This is certainly a drawback in
the current study, but it was chosen as a compromise in order to be
able to explore
a larger parameter space with given computational resources.

Contrary to the flame evolution in Fig.~\ref{evo_vort_fig}a, the case of
$v'/s_l = 2.5$ 
(Fig.~\ref{evo_vort_fig}d) clearly shows the disruption of the initial
cellular pattern by interaction of the flame with vortices. Also, the
formation of a single domain-filling cell is suppressed by the
interaction. 
The overall flame shape evolution in this case can be interpreted as
an adaptation to the imprinted vortical flow structure.
This is emphasized by Fig~\ref{panel_fig_3}. It provides a more
detailed visualization of the 
interaction between the flame shape and the vortical flow, which is
characterized by the color-coded
logarithm of the absolute value of the
vorticity $\zeta$ of the flow field,
\begin{equation}\label{vorticity_def}
  \zeta = \frac{\partial v_y}{\partial x} - \frac{\partial
    v_x}{\partial y},
\end{equation}
pointing out the vortices imprinted on the fuel flow.

The transition in the flame evolution between the  extreme cases
depicted in Figs.~\ref{evo_vort_fig}a and \ref{evo_vort_fig}d does,
however, not proceed abruptly, but rather in a smooth
transition. Figures \ref{evo_vort_fig}b,c present two examples of
intermediate behaviors at $v'/s_l = 0.7$ and $v'/s_l = 2.0$, the first
of which is again illustrated by snapshots in Fig.~\ref{panel_fig_2}.
In this case, still a tendency to form a long-wavelength cellular
structure is visible. Here the mechanism of the advection of
small perturbations toward the cusp \citep{zeldovich1980a, roepke2003a,
  roepke2003b} dominates over the effects of the vortical flow on the
flame shape.

Simulations similar to the aforementioned were also performed in some cases with higher
resolution of the numerical grid in order to test whether or not insufficient
resolution prevents small scale flame structures from
developing. However, no significant difference to the flame evolution
depicted in Fig.~\ref{evo_vort_fig}a--d could be noticed. 
This agrees with \citet{helenbrook1999a}, who found in their
simulations of chemical flames that the wavelength of perturbations
that develop in the interaction with a vortical flow is determined by the
scale of the vortices and not by a critical wavelength (introduced by
a curvature-dependent burning speed of the flame front according to
\citealt{markstein1951a}).

\subsection{A Parameter study}
\label{param_sec}

In order to quantify the flame evolution as
qualitatively described in the preceding section, we conducted a
parameter study. It aimed at the determination of
flame behavior as a function of the strength of the
vortices in the incoming flow and in dependence on the fuel
density. 

As discussed by \citet{roepke2003b}, the simulation of flame
stabilization at around $1.0 \times 10^7
\,\mathrm{g} \,\mathrm{cm}^{-3}$ requires unaffordable high numerical
resolutions and furthermore our thin flame approximation is certainly not a
realistic description of thermonuclear burning in SNe Ia at these low
densities. On the other hand, it can be
observed, that at higher fuel densities (around $10^8 \,\mathrm{g}
\,\mathrm{cm}^{-3}$) the sensitivity of the initial 
flame to small-wavelength perturbations increases. This causes
difficulties in producing a stable flame configuration prior to the
incoming vortices encountering the flame front. However, in
connection with a possible DDT we are mainly interested in the late
stages of the SN Ia explosions (see Sect.~\ref{intro_sec}). Therefore
we restricted the study to four fuel densities: $\rho_u = 1.25 \times
10^7 \,\mathrm{g} \,\mathrm{cm}^{-3}$, $\rho_u = 2.5 \times 10^7 \,\mathrm{g}
\,\mathrm{cm}^{-3}$, $\rho_u = 5 \times 10^7 \,\mathrm{g} \,\mathrm{cm}^{-3}$, and
$\rho_u = 7.5 \times 10^7 \,\mathrm{g} \,\mathrm{cm}^{-3}$ (cf.~Table
\ref{values_tab}). 

Plots similar to Fig.~\ref{evo_vort_fig}
corresponding to simulations with 
different fuel densities are given by \cite{roepke2003diss}.
The flame evolution in these simulations can be summarized as follows:
At
lower fuel densities the effect of the incoming vortices on the
flame structure is generally more drastic. For $\rho_u=2.5 \times 10^7 \,\mathrm{g}
\,\mathrm{cm}^{-3}$ the tendency of the formation of
the domain-filling structure is still visible in the case of
propagation into quiescent fuel. This is not the case for
$\rho_u=1.25 \times 10^7 \,\mathrm{g} \,\mathrm{cm}^{-3}$
for reasons discussed by \citet{roepke2003b}. However, with increasing strengths of the
imprinted velocity fluctuations, the flame still gradually adapts to
the flow for both fuel densities.
At $\rho_u = 7.5 \times 10^7 \,\mathrm{g} \,\mathrm{cm}^{-3}$ the flame evolution
is similar to that at
$\rho_u= 5 \times 10^7 \,\mathrm{g} \,\mathrm{cm}^{-3}$. Only in case of
propagation into quiescent fuel an initial
destabilization with respect to small wavelength perturbations
(cf.~\citealt{roepke2003b}) alters the evolution for a transition
period, before the flame finally stabilizes in a domain-filling single-cell
structure.

In a quantitative evaluation of our parameter study regarding the flame
evolution for different fuel densities we will now address
(i) the dependency of the effective flame propagation velocity on the strength
of the imprinted velocity fluctuations and (ii) the amplification of
the velocity fluctuation by the flame front. 

For the exemplary case of $\rho_u= 5 \times 10^7 \,\mathrm{g} \,\mathrm{cm}^{-3}$
we discuss the measurement of the necessary quantities in our
simulations. 
\begin{figure}[t]
\centerline{\resizebox{0.9 \hsize}{!}{
\includegraphics[viewport = 0 0 251 506]{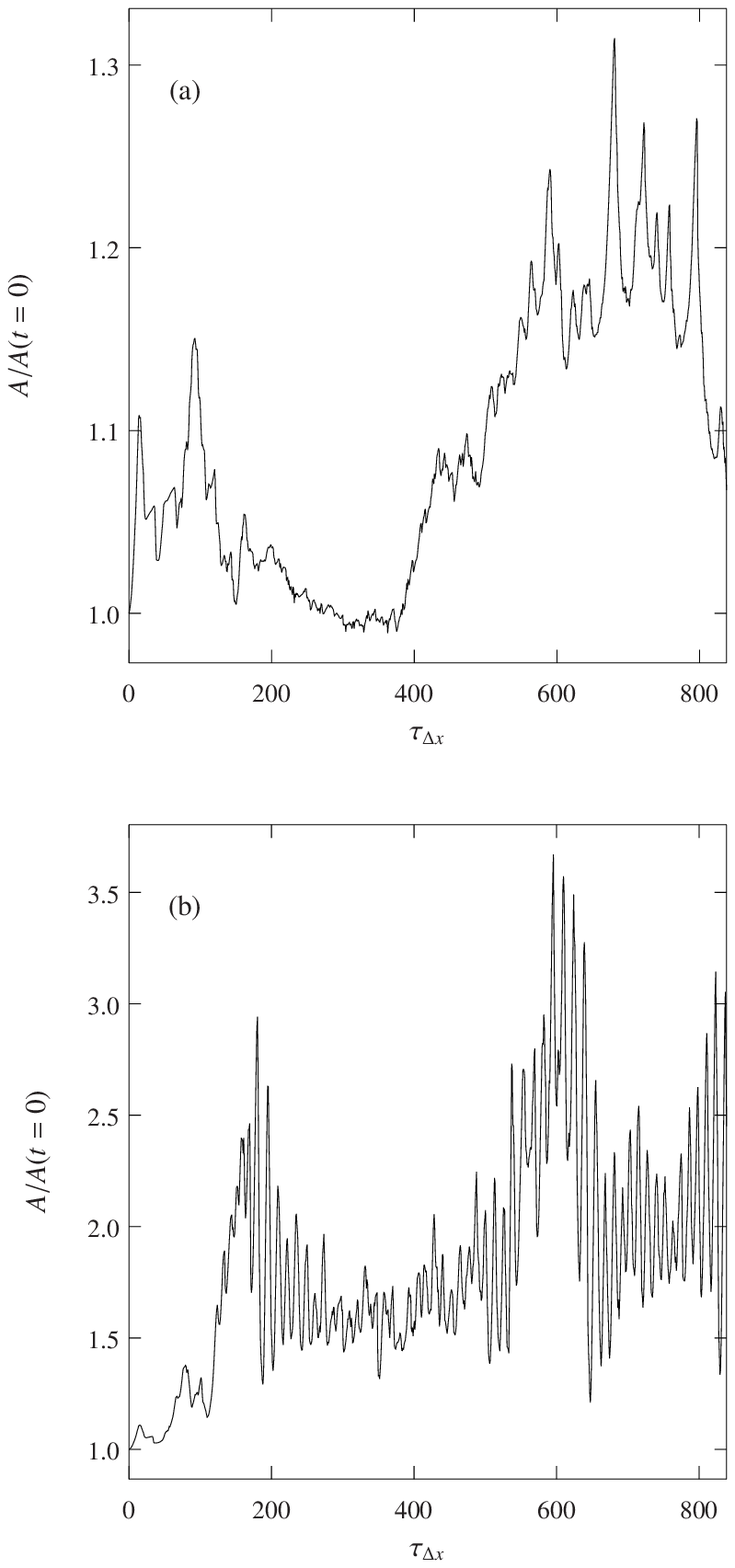}}}
\caption{Flame surface area over time (normalized to the crossing time of the
  laminar flame over one grid cell $\tau_{\Delta x}$) for
  \textbf{(a)} $v'/s_l$ = 0.7 and \textbf{(b)} $v'/s_l = 2.5$.\label{area_vorf_fig}}
\end{figure}
\begin{figure*}[t]
\centerline{\includegraphics[width=0.8 \hsize, viewport=0 0 323 310,clip]{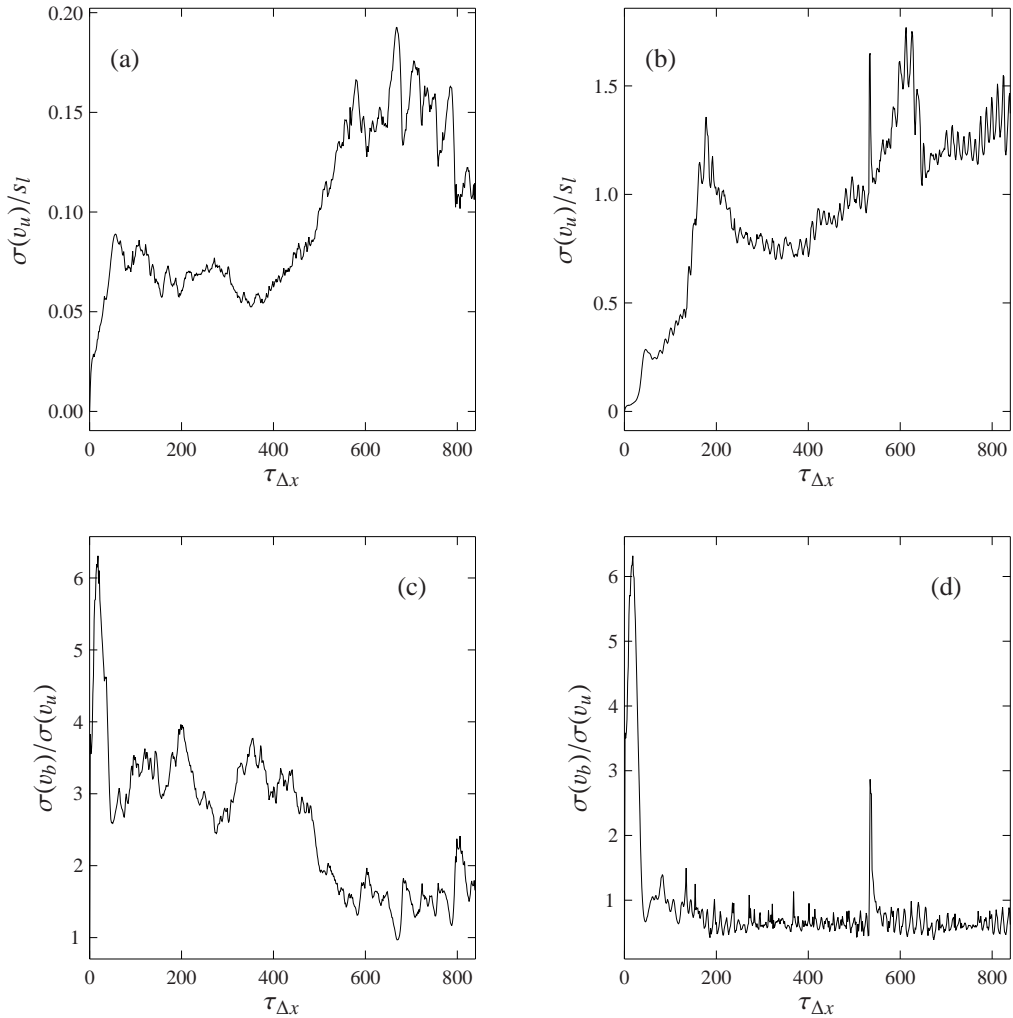}}
\caption{\textbf{(a,b)} standard deviation of the velocity field ahead
  of the front over time, and \textbf{(c,d)} ratio of the standard
  deviations of the velocity fields beyond and ahead of the flame over
  time. The left and
  right columns of plots correspond to $v'/s_l$ = 0.7 and $v'/s_l =
  2.5$, respectively.\label{standard_fig}}
\end{figure*}
The effective propagation velocity $v_\mathrm{eff}$ of the flame is determined via the
flame surface:
\begin{equation}\label{v_eff_eq}
\frac{v_\mathrm{eff}}{s_l} = \frac{A_\mathrm{surf}}{A_\mathrm{planar}}.
\end{equation}
Underlying to that relation is the assumption of 
a constant, geometry independent burning velocity $s_l$ of the
flame. This assumption has been discussed in the context of
the present flame model by \citet{roepke2003a}.
The measured flame surface area as a function of time is plotted in
Figs.~\ref{area_vorf_fig}a and \ref{area_vorf_fig}b for the simulations with $v'/s_l=0.7$ and
$v'/s_l=2.5$, respectively. Here adaptation of the flame structure to
the strong vortical fuel flow is obvious. After the imprinted velocity
fluctuations reach the flame at $t\sim 150 \tau_{\Delta x}$, its
surface becomes dominated by the vortices and changes with the
frequency of these. This is reflected by 
the high-frequency fluctuations of $A$ in Fig.~\ref{area_vorf_fig}b. 
In addition to fluctuations on a small time
scale, the interaction with the vortical flow field introduces
considerable long term fluctuations. These are, of course, expected in
the initial phase when the flame begins to react to the incoming
vortices. But even at much later times the flame surface area
does not reach a steady state.

The standard deviation of the velocity field is obtained separately in
the fuel upstream of the front and in the ashes downstream of it in
order to determine the amplification of the velocity fluctuations
across the flame front. In Fig.~\ref{standard_fig} the quantities are plotted
against time for our exemplary case. 
As mentioned above, the strength of the
vortices produced at the inflow boundary is not a reliable measure of
what the flame actually experiences, since the velocity fluctuations
get damped quickly when propagating toward the flame and bending of
the flame front may lead to different velocity fluctuations at
different locations on the flame. Moreover, the dependence of the damping
on the fuel density hampers the comparability of simulations with this
parameter varying. Therefore we determine the standard deviations
of the velocity fields, $\sigma(v_u)$ and $\sigma(v_b)$, within a certain belt
around the flame front. This belt was
scaled in a way that the time for flame crossing over its width remained
the same for all fuel densities.

All quantities measured in our simulations
fluctuate considerably with time. Hence we averaged them over a
time ranging from 480 $\tau_{\Delta x}$ to 720 $\tau_{\Delta
  x}$ ($\tau_{\Delta x}$ denotes the crossing time of the
corresponding laminar flame over one grid cell). 

\begin{figure}[t]
\centerline{\resizebox{\hsize}{!}{
\includegraphics{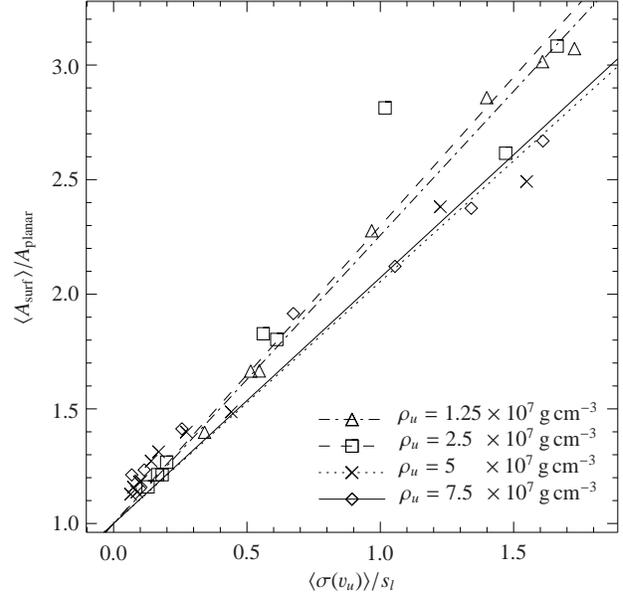}}}
\caption{Dependence of the flame surface area on the strength of the
  imprinted velocity fluctuations.\label{fit_1_fig}}
\end{figure}

In Fig.~\ref{fit_1_fig} the temporal mean value of flame surface
area $A_\mathrm{surf}$ (normalized to the surface area of the corresponding planar flame
front $A_\mathrm{planar}$) is plotted against the temporal mean of standard deviation of
the velocity in the fuel region (normalized to the laminar burning
velocity of the flame).
The increase of the flame speed with stronger incoming velocity
fluctuations is evident here. However, there is
quite a large scatter in the data. This has two reasons. First, we
certainly did not follow the flame evolution long enough to
undoubtly reach the steady state of the flame shape---which should
ideally be the clearly distinct single domain-filling cusp-like
structure in case of weak incoming vortices. To follow the flame
propagation up to this stage would, however, be much too expensive for
a parameter study. Second, some peculiar features of the flame shape may
develop as a result of small perturbations (e.g.~due to numerical
noise). 
In particular, this may cause substantial deviations if it affects the
long-wavelength structure of the flame and is primarily
responsible for the
scatter of the results in case of strong incoming velocity
fluctuations. This scatter could very likely be cured
by taking the mean over longer time intervals.

\begin{figure}[t]
\centerline{\resizebox{\hsize}{!}{
\includegraphics{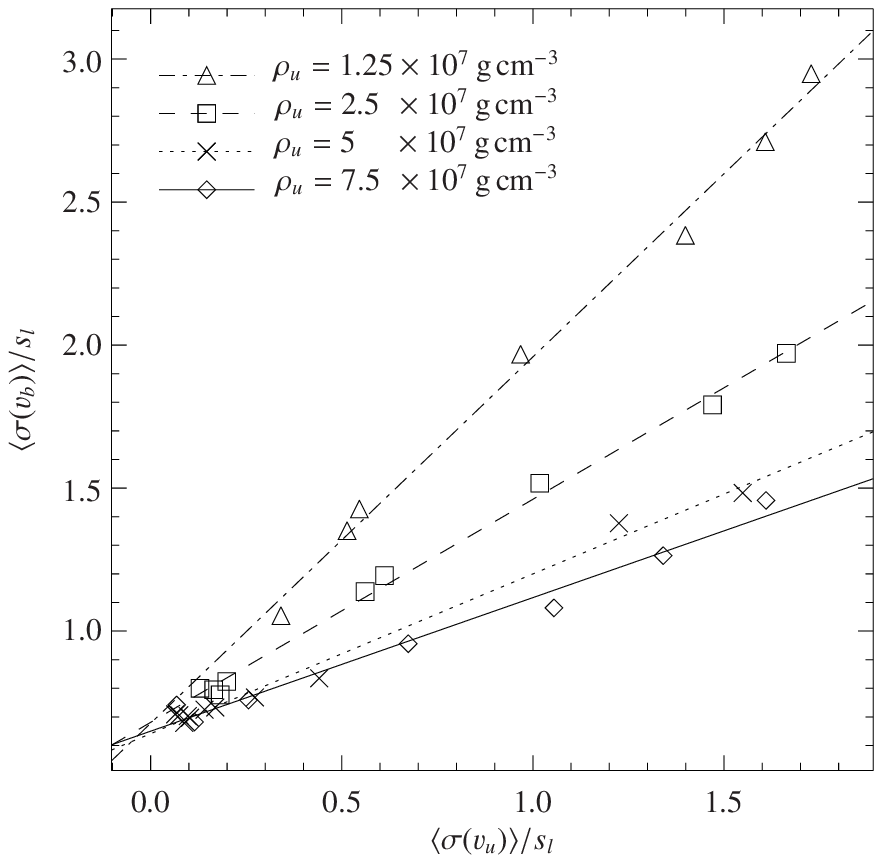}}}
\caption{Relation between velocity fluctuations upstream and
  downstream of the flame front.\label{fit_2_fig}}
\end{figure}

\citet{damkoehler1940a} proposed a linear dependence of the
effective flame propagation velocity on the turbulence intensity
for turbulent combustion in what
is today called the flamelet regime \citep{peters1986a}. Our results are consistent
with a linear growth law.
Corresponding fits to the data are included in
Fig.~\ref{fit_1_fig} and values for the parameter $a$ according to the fitting formula
\begin{equation}\label{fit_form_1}
\frac{\langle A_\mathrm{surf}\rangle}{A_\mathrm{planar}} = 1.0 + a \,
\frac{\langle\sigma(v_u)\rangle}{s_l}
\end{equation}
are given in Table \ref{fit_tab}. It is evident from Fig.~\ref{fit_1_fig}
that interaction of the flame in the cellular regime with turbulent
velocity fluctuation can lead to
a substantial acceleration of the flame propagation velocity.
\begin{table*}[ht]
\caption{Fit parameters according to fitting formulas \req{fit_form_1}
  and \req{fit_form_2} with corresponding asymptotic standard errors.
\label{fit_tab}}
$$
\begin{array}{p{0.11\linewidth}p{0.16\linewidth}p{0.16\linewidth}p{0.16\linewidth}}
\hline\hline
\noalign{\smallskip}
\mbox{$\rho_u$ [$\,\mathrm{g} \,\mathrm{cm}^{-3}$]} & \mbox{$ a $} & \mbox{$b$} & \mbox{$c$} \\
\hline
\mbox{$1.25 \times 10^7$} & \mbox{$1.25712 \pm 0.0215$}  & \mbox{$ 0.177654 \pm 0.05236$} &
\mbox{$ 1.28103 \pm 0.0.04581$}\\
\mbox{$2.5 \hspace{0.5em}\times 10^7$}  & \mbox{$1.29816 \pm 0.0797$}  & \mbox{$ 0.180935 \pm 0.01718$} & \mbox{$0.779895 \pm 0.01982$}\\
\mbox{$5 \hspace{1.25em} \times 10^7$}  & \mbox{$1.05533 \pm 0.04931$} & \mbox{$
0.141428 \pm 0.01171$} & \mbox{$0.558411 \pm 0.01883$}\\
\mbox{$7.5 \hspace{0.5em}\times 10^7$}  & \mbox{$1.07245 \pm 0.05131$} & \mbox{$ 0.150079 \pm 0.2911$} &  \mbox{$0.466975 \pm 0.03134$}\\
\noalign{\smallskip}
\hline
\end{array}
$$
\end{table*}
Unfortunately, it is not possible  to infer a clear
trend of the slope depending on the fuel density from this parameter study.

In order to estimate the amplification of the velocity fluctuations
across the flame for different densities, we compared the
strength of velocity fluctuations downstream of the flame with the
strength of the imprinted vortices in the fuel. As already shown
for the case of flame propagation into quiescent fuel
\citep{roepke2003b}, the 
flame produces vorticity in the ashes. Here, the question is addressed whether
vorticity present in the fuel will be amplified in the flame.

The result of the
study for a variety of fuel densities is plotted in Figure
\ref{fit_2_fig}. In this plot,
trends show up much more
clearly than in Fig.~\ref{fit_1_fig}. However, some scatter is still present in the
data. As can be
seen from snapshots of the evolution of the flame front (an example is
the bottom left snapshot in Fig.~\ref{panel_fig_3}), the merging
of cusps can eventually produce
transient ``bursts'' in the vorticity downstream of the flame. This is
particularly prominent in case of changes in the long-wavelength flame
structure, but not necessarily connected to it. From Fig.~\ref{standard_fig}d, where this ``burst'' in velocity fluctuation
appears as a spike in the profile, it can be concluded that the duration of those events is very
short. Hence their contribution
to the temporal mean is small (whereas the long-wavelength flame
structure and thus the flame surface changes slowly). Consequently, the
scatter in plot \ref{fit_2_fig} is much smaller than in plot \ref{fit_1_fig}.

\begin{figure}[t!]
\centerline{\resizebox{\hsize}{!}{
\includegraphics{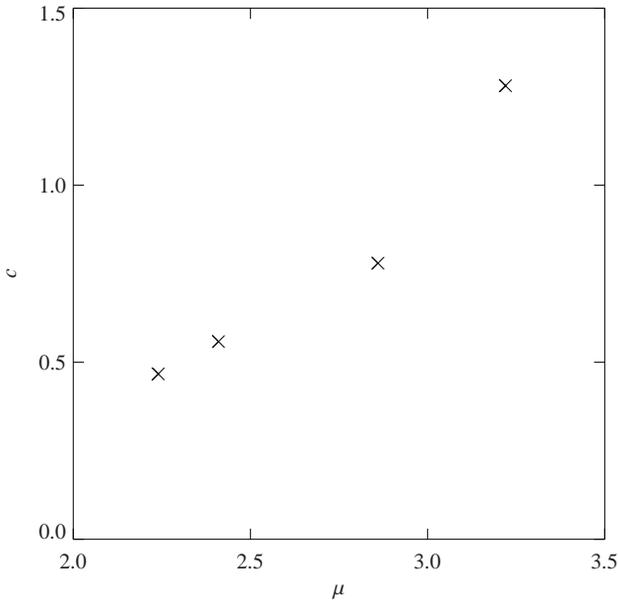}}}
\caption{Ratio of the velocities upstream and downstream of the flame
  front as a function of the density contrast $\mu$ over it.\label{muamp_fig}}
\end{figure}

The data plotted in Fig.~\ref{fit_2_fig} can be fit rather well by a linear
relation between the ratio $\langle\sigma (v_b)/
\sigma(v_u)\rangle$ and the incoming turbulence
intensity, i.e.
\begin{equation}\label{fit_form_2}
\langle\sigma(v_b)\rangle = b + c \langle\sigma(v_u)\rangle.
\end{equation}
Table \ref{fit_tab} provides the fit parameters. 
In contrast to Figure \ref{fit_1_fig}, the plot in Figure
\ref{fit_2_fig} reveals a clear trend for the fuel density. 
The slope $c$ increases with lower $\rho_u$. This is what
would be expected taking into account the
increased production of specific volume across the flame with lower $\rho_u$. In
Fig.~\ref{muamp_fig} the factor $c$ is plotted against the corresponding
density ratio $\mu$, accentuating the trend. Of course, a
functional dependence cannot be inferred from this small sample of data.
It is interesting to note  that for
$\rho_u \approx 2.5 \times 10 ^7 \,\mathrm{g} \,\mathrm{cm}^{-3}$ the slope $c$ becomes
greater than unity, indicating a slight amplification of velocity
fluctuation in this case.

The above interpretation of this result requires some
caution, since the cellular stabilization of the flame has been shown
to be weaker with lower fuel density. However, this
seems to be a numerical rather than a physical effect, since the cusps
become more stable with higher resolution. Thus, at the given
resolution of our parameter study, the flame evolution shows peculiar
behavior below a fuel density of $1.25 \times 10^7 \,\mathrm{g}
\,\mathrm{cm}^{-3}$. Here, cusps may loose stability distorting the flame
shape, which then responds with accelerated propagation of the
perturbed part for a short period, after that the flame stabilizes
again.
However, we do
not observe such a flame evolution at higher fuel densities included
in this parameter study. Thus, the dependency on the fuel density is likely
to be (at least partially) of physical origin.

\section{Conclusions}

In the presented simulations we performed the first attempt to
investigate the interaction of cellular flames with turbulent velocity
fluctuations in the context of SN Ia explosions. The results can be
summarized as follows:
\begin{enumerate}
\item
In case of the interaction of the flame with weak vortices in the fuel
flow, the flame exhibits an evolution similar to that observed in case
of flame propagation into quiescent fuel. In the setup chosen for our
simulations the flame front showed a tendency to align in a
single domain-filling cellular structure. Interaction with the
incoming vortices leads to a
superposition of this fundamental structure with a cellular pattern of
shorter wavelength. Similar to the simulations presented by
\citet{roepke2003a}, where a cellular superposition of the basic flame
structure was observed in case of wide computational domains, this
does not lead to a destabilization of the domain-filling cell. In
accord with theoretical predictions \citep{zeldovich1980a} the small
superimposed perturbations are advected toward the cusp where they
disappear. The flow field accounting for this effect was studied by
\citet{roepke2003b}.
\item
If the cellularly stabilized flame interacts with vortices of
sufficient strength, then the initial cellular pattern will
break up in the sense that the initial shape is lost and the flame
adapts to the incoming vortical flow. The tendency of the flame to
form a single-cell structure is suppressed by this effect. However,
we do not observe a sharp transition but rather a gradual change in the
flame behavior.
\item
Our numerical simulations suggest that the flame smoothly adapts
to the incoming vortical flow with increasing turbulence
intensity. No drastic effects were observed. Although we measured
an amplification of the strength of the velocity fluctuation across
the flame for $\rho_u=2.5\times 10^7 \,\mathrm{g} \,\mathrm{cm}^{-3}$, we could
not observe a deviation from a linear scaling between the
flame surface area
(and thus the effective flame propagation speed) and the strength of
the imprinted velocity fluctuations.
\end{enumerate}

We draw the conclusion that the results of our study are consistent
with flame propagation in the flamelet
regime of turbulent combustion,  at least at fuel
densities above $\rho_u \approx 2.5\times 10^7 \,\mathrm{g}
\,\mathrm{cm}^{-3}$.
Thus, our study corroborates the assumption of flame
stability at unresolved scales in large-scale SN Ia explosion
models. This is an important contribution that our small-scale model
makes to the credibility of those simulations. The slight amplification of the turbulent
velocity fluctuations across the flame at $\rho_u =  2.5\times 10^7 \,\mathrm{g}
\,\mathrm{cm}^{-3}$ (and possibly even more pronounced below that value) may
have no significant impact on the current SN Ia models, since the flame
width at those densities will become non-negligible (cf.~the
data given by \citealt{timmes1992a}). Then the burning is expected to
enter the distributed regime of turbulent combustion, whose simulation requires
completely different numerical methods. In the context of SN Ia
explosions, an approach to study this regime was discussed by
\citet{lisewski2000a}.

Although the increase in effective flame propagation resulting from
the cellular regime is negligible compared to the increase in
flame speed at larger scales where the flame is affected by the
turbulent cascade, it may be significant in the early stages of the
explosion. In the current large-scale models (e.g.~\citealt{reinecke2002d}), the flame is assumed to
propagate with its laminar flame velocity before the rising Rayleigh-Taylor bubbles
establish the turbulent cascade. This may cause a problem with the
nucleosynthetic yields from the SN Ia explosion \citep{reinecke_phd}. Since the laminar
flame speed is very low, the WD expands slowly in
the beginning of the explosion and thus the combustion products remain
at rather high densities for a considerable time. This effect causes
a neutronization of the material by electron capture and an
overproduction of neutron-rich heavy nuclei
\citep{nomoto1991a,brachwitz2000a}, eventually even leading to a collapse
of the star. Taking into account the velocity increase resulting from
the cellular regime, this problem could be extenuated. As has been
shown in this study, this may be in
particular the case if strong velocity fluctuations left over from
the pre-ignition convection interact with the flame. However, the
strength of turbulence resulting from pre-ignition effects is not
well-determined yet. \citet{hoeflich2002a} claim values as high as
$\sim$$10^7 \,\mathrm{cm} \,\mathrm{s}^{-1}$. The impact of the cellular
burning regime on the nucleosynthesis yields of the SN Ia explosion
models will be investigated in a forthcoming study. 

Even though we can probably not completely rule out the possibility of
active turbulent combustion, we found no convincing hint for such an effect in
our numerical investigations.
Our simulations indicate that effects
resulting from the cellular 
regime of flame propagation are unlikely to trigger a presumed
deflagration-to-detonation transition. Hence, the search for active turbulent
combustion and deflagration-to-detonation transition should focus on
the distributed burning regime.

\begin{acknowledgements}
This work was supported in part by the European Research Training
Network ``The Physics of Type Ia Supernova Explosions'' under contract
HPRN-CT-2002-00303 and by the DFG Priority Research Program ``Analysis
and Numerics for Conservation Laws'' under contract HI 534/3.
A pleasant atmosphere to prepare this publication was provided at the
workshop ``Thermonuclear Supernovae and Cosmology'' at the ECT*,
Trento, Italy. 
We would like to thank M.~Reinecke,
S.~Blinnikov, and W.~Schmidt for stimulating discussions.
The numerical simulations were performed on an IBM Regatta
system at the computer center of the Max Planck Society in Garching.
\end{acknowledgements}


\begin{thebibliography}{41}
\expandafter\ifx\csname natexlab\endcsname\relax\def\natexlab#1{#1}\fi

\bibitem[{{Arnett}(1969)}]{arnett1969a}
{Arnett}, W.~D. 1969, Ap\&SS., 5, 180

\bibitem[{{Blinnikov} \& {Sasorov}(1996)}]{blinnikov1996a}
{Blinnikov}, S.~I. \& {Sasorov}, P.~V. 1996, Phys. Rev. E, 53, 4827

\bibitem[{{Brachwitz} {et~al.}(2000){Brachwitz}, {Dean}, {Hix}, {Iwamoto},
  {Langanke}, {Mart{\'{\i}}nez-Pinedo}, {Nomoto}, {Strayer}, {Thielemann}, \&
  {Umeda}}]{brachwitz2000a}
{Brachwitz}, F., {Dean}, D.~J., {Hix}, W.~R., {et~al.} 2000, ApJ,
  536, 934

\bibitem[{{Bychkov} \& {Liberman}(1995)}]{bychkov1995a}
{Bychkov}, V. \& {Liberman}, M.~A. 1995, A\&A, 302, 727

\bibitem[{{Colella} \& {Woodward}(1984)}]{colella1984a}
{Colella}, P. \& {Woodward}, P.~R. 1984, J. Comput. Phys., 54, 174

\bibitem[{{Damk{\"o}hler}(1940)}]{damkoehler1940a}
{Damk{\"o}hler}, G. 1940, Z. f. Elektroch., 46, 601

\bibitem[{{Darrieus}(1938)}]{darrieus1938a}
{Darrieus}, G. 1938, communication presented at \emph{La Technique Moderne},
  Unpublished.

\bibitem[{{Fryxell} {et~al.}(1989){Fryxell}, {M{\"u}ller}, \&
  {Arnett}}]{fryxell1989a}
{Fryxell}, B.~A., {M{\"u}ller}, E., \& {Arnett}, W.~D. 1989, Hydro\-dynamics and
  nuclear burning, MPA Green Report 449, Max-Planck-Institut f\"ur Astrophysik,
  Garching

\bibitem[{{Gamezo} {et~al.}(2003){Gamezo}, {Khokhlov}, {Oran}, {Chtchelkanova},
  \& {Rosenberg}}]{gamezo2003a}
{Gamezo}, V.~N., {Khokhlov}, A.~M., {Oran}, E.~S., {Chtchelkanova}, A.~Y., \&
  {Rosenberg}, R.~O. 2003, Science, 299, 77

\bibitem[{{Gutman} \& {Sivashinsky}(1990)}]{gutman1990a}
{Gutman}, S. \& {Sivashinsky}, G.~I. 1990, Physica D, 43, 129

\bibitem[{{Helenbrook} \& {Law}(1999)}]{helenbrook1999a}
{Helenbrook}, B.~T. \& {Law}, C.~K. 1999, Combustion and Flame, 117, 155

\bibitem[{{Hillebrandt} \& {Niemeyer}(2000)}]{hillebrandt2000a}
{Hillebrandt}, W. \& {Niemeyer}, J.~C. 2000, ARA\&A, 38,
  191

\bibitem[{{H\"oflich} \& {Khokhlov}(1996)}]{hoeflich1996a}
{H\"oflich}, P. \& {Khokhlov}, A. 1996, ApJ, 457, 500

\bibitem[{{H\"oflich} \& {Stein}(2002)}]{hoeflich2002a}
{H\"oflich}, P. \& {Stein}, J. 2002, ApJ, 568, 779

\bibitem[{Hoyle \& Fowler(1960)}]{hoyle1960a}
Hoyle, F. \& Fowler, W.~A. 1960, ApJ, 132, 565

\bibitem[{{Iwamoto} {et~al.}(1999){Iwamoto}, {Brachwitz}, {Nomoto},
  {Kishimoto}, {Umeda}, {Hix}, \& {Thielemann}}]{iwamoto1999a}
{Iwamoto}, K., {Brachwitz}, F., {Nomoto}, K., {et~al.} 1999, ApJS, 125, 439

\bibitem[{Kerstein(1996)}]{kerstein1996a}
Kerstein, A.~R. 1996, Combust. Sci. Technol., 118, 189

\bibitem[{{Landau}(1944)}]{landau1944a}
{Landau}, L.~D. 1944, Acta Physicochim. URSS, 19, 77

\bibitem[{{Lisewski} {et~al.}(2000){Lisewski}, {Hillebrandt}, {Woosley},
  {Niemeyer}, \& {Kerstein}}]{lisewski2000a}
{Lisewski}, A.~M., {Hillebrandt}, W., {Woosley}, S.~E., {Niemeyer}, J.~C., \&
  {Kerstein}, A.~R. 2000, ApJ, 537, 405

\bibitem[{{Markstein}(1951)}]{markstein1951a}
{Markstein}, G.~H. 1951, J.~Aeronaut.~Sci., 18, 199

\bibitem[{{Niemeyer}(1999)}]{niemeyer1999a}
{Niemeyer}, J.~C. 1999, ApJ, 523, L57

\bibitem[{{Niemeyer} \& {Hillebrandt}(1995)}]{niemeyer1995a}
{Niemeyer}, J.~C. \& {Hillebrandt}, W. 1995, ApJ, 452, 779

\bibitem[{{Niemeyer} \& {Woosley}(1997)}]{niemeyer1997b}
{Niemeyer}, J.~C. \& {Woosley}, S.~E. 1997, ApJ, 475, 740

\bibitem[{{Nomoto} \& {Kondo}(1991)}]{nomoto1991a}
{Nomoto}, K. \& {Kondo}, Y. 1991, ApJ, 367, L19

\bibitem[{{Osher} \& {Sethian}(1988)}]{osher1988a}
{Osher}, S. \& {Sethian}, J.~A. 1988, J. Comput. Phys., 79, 12

\bibitem[{{Peters}(1986)}]{peters1986a}
{Peters}, N. 1986, in Twenty-First Symposium (International) on Combustion
  (Pittsburgh: The Combustion Institute), 1231--1250

\bibitem[{{Reinecke} {et~al.}(2002){Reinecke}, {Hillebrandt}, \&
  {Niemeyer}}]{reinecke2002d}
{Reinecke}, M., {Hillebrandt}, W., \& {Niemeyer}, J.~C. 2002, A\&A, 391, 1167

\bibitem[{{Reinecke} {et~al.}(1999){Reinecke}, {Hillebrandt}, {Niemeyer},
  {Klein}, \& {Gr{\" o}bl}}]{reinecke1999a}
{Reinecke}, M., {Hillebrandt}, W., {Niemeyer}, J.~C., {Klein}, R., \&
  {Gr{\"o}bl}, A. 1999, A\&A, 347, 724

\bibitem[{{Reinecke}(2001)}]{reinecke_phd}
{Reinecke}, M.~A. 2001, PhD thesis, Technical University of Munich

\bibitem[{{Richardson}(1922)}]{richardson1922a}
{Richardson}, L.~F. 1922, Weather prediction by numerical process (Cambridge:
  Cambridge University Press), (republished Dover 1965)

\bibitem[{{R{\"o}pke}(2003)}]{roepke2003diss}
{R{\"o}pke}, F.~K. 2003, PhD thesis, Technical University of Munich, available
  at http://tumb1.ub.tum.de/publ/diss/

\bibitem[{{R{\"o}pke} {et~al.}(2003{\natexlab{a}}){R{\"o}pke}, {Niemeyer}, \&
  {Hillebrandt}}]{roepke2003b}
{R{\"o}pke}, F.~K., {Niemeyer}, J.~C., \& {Hillebrandt}, W. 2003{\natexlab{a}},
  submitted to A\&A

\bibitem[{{R{\"o}pke} {et~al.}(2003{\natexlab{b}}){R{\"o}pke}, {Niemeyer}, \&
  {Hillebrandt}}]{roepke2003a}
{R{\"o}pke}, F.~K., {Niemeyer}, J.~C., \& {Hillebrandt}, W. 2003{\natexlab{b}},
  ApJ, 588, 952

\bibitem[{{Smiljanovski} {et~al.}(1997){Smiljanovski}, {Moser}, \&
  {Klein}}]{smiljanovski1997a}
{Smiljanovski}, V., {Moser}, V., \& {Klein}, R. 1997, Combustion Theory and
  Modelling, 1, 183

\bibitem[{{Sussman} {et~al.}(1994){Sussman}, {Smereka}, \&
  {Osher}}]{sussman1994a}
{Sussman}, M., {Smereka}, P., \& {Osher}, S. 1994, J. Comput. Phys., 114, 146

\bibitem[{Thual {et~al.}(1985)Thual, Frisch, \& H\'enon}]{thual1985a}
Thual, O., Frisch, U., \& H\'enon, M. 1985, J.~Phys.~(Paris), 46, 1485

\bibitem[{{Timmes} \& {Woosley}(1992)}]{timmes1992a}
{Timmes}, F.~X. \& {Woosley}, S.~E. 1992, ApJ, 396

\bibitem[{{Vladimirova} {et~al.}(2003){Vladimirova}, {Constantin}, {Kiselev},
  {Ruchayskiy}, \& {Ryzhik}}]{vladimirova2003a}
{Vladimirova}, N., {Constantin}, P., {Kiselev}, A., {Ruchayskiy}, O., \&
  {Ryzhik}, L. 2003, Combustion Theory and Modelling, (in press) Preprint
  available: physics/0212057

\bibitem[{{Zel'dovich}(1966)}]{zeldovich1966a}
{Zel'dovich}, Y.~B. 1966, Journal of Appl. Mech. and Tech. Physics, 1, 68,
  english translation

\bibitem[{{Zel'dovich} {et~al.}(1980){Zel'dovich}, {Istratov}, {Kidin}, \&
  {Librovich}}]{zeldovich1980a}
{Zel'dovich}, Y.~B., {Istratov}, A.~G., {Kidin}, N.~I., \& {Librovich}, V.~B.
  1980, Combust. Sci. Technol., 24, 1

\bibitem[{{Zhu} \& {Ronney}(1994)}]{zhu1994a}
{Zhu}, J. \& {Ronney}, P.~D. 1994, Combust. Sci. Technolog., 100, 183

\end{thebibliography}
\end{document}